\documentclass[twocolumn,showpacs,preprintnumbers,amsmath,amssymb,superscriptaddress]{revtex4}
\usepackage{amsmath}    
\usepackage{graphicx}   

\usepackage{dcolumn,bm,amsmath}

\newcommand{\TwoByTwo}[4]{\mbox{$\left(\begin{array}{cc}#1 & #2 \\ #3 & #4 \end{array}\right)$}}
\newcommand{\TwoColumn}[2]{\mbox{$\left(\begin{array}{cc}#1 \\ #2 \end{array}\right)$}}

\DeclareMathAlphabet\EuScript{U}{eus}{m}{n}
\SetMathAlphabet\EuScript{bold}{U}{eus}{b}{n}
\newcommand{\Bvec}{\boldsymbol{\EuScript{B}}}%
\newcommand{\Bsca}{\EuScript{B}}

\begin{document}

\title{Using Molecules to Measure Nuclear Spin-Dependent Parity Violation}
\author{D. DeMille, S.B. Cahn, D.Murphree, D.A.Rahmlow}
\affiliation{Department of Physics, P.O. Box 208120, Yale University, New Haven, Connecticut 06520}
\author{M.G. Kozlov}
\affiliation{Petersburg Nuclear Physics Institute, Gatchina, 188300, Russia}

\begin{abstract}
Nuclear spin-dependent parity violation arises from weak interactions between electrons and nucleons, and from nuclear anapole moments. We outline a method to measure such effects, using a Stark-interference technique to determine the mixing between opposite-parity rotational/hyperfine levels of ground-state molecules. The technique is applicable to nuclei over a wide range of atomic number, in diatomic species that are theoretically tractable for interpretation. This should provide data on anapole moments of many nuclei, and on previously unmeasured neutral weak couplings.
\end{abstract}


\maketitle


Up to now, atomic parity violation (PV) experiments have primarily focused on the PV effect arising from the weak charge of the nucleus $Q_{\rm W}$ \cite{Reviews}, a nuclear spin-independent quantity that parameterizes the electroweak neutral coupling between electron axial- and nucleon vector-currents ($A_eV_n$).  Here we propose a highly sensitive and widely applicable technique to measure nuclear spin-\emph{dependent} (NSD) PV effects.  Such effects arise primarily from two underlying causes.  The first is the nuclear anapole moment, a P-odd magnetic moment induced by weak interactions within the nucleus, which couples to the spin of a penetrating electron \cite{Flambaum80}.  Measurements of anapole moments can provide useful data on purely hadronic PV interactions \cite{FlamAnapoleDDH,HaxWie}.  So far, only one nuclear anapole moment has been measured, in $^{133}$Cs \cite{Wood97}.  The second source of NSD-PV is the electroweak neutral coupling between electron vector- and nucleon axial-currents ($V_eA_n$).  This can be parameterized by two constants, $C_{2{\rm u,d}}$, which describe the $V_eA_n$ couplings to up and down quarks.  These are suppressed in the Standard Model (SM), making $C_{2{\rm u,d}}$ difficult to measure and at present perhaps the most poorly characterized parameters in the SM \cite{PDG06}.  However, because of this suppression, even moderately precise measurements of $C_{2{\rm u,d}}$ could be sensitive to new physics at TeV energy scales \cite{Langacker92}.

Our method to measure NSD-PV exploits the properties of diatomic molecules \cite{SF,FK85,KL} to amplify the observable signals.  Rotational/hyperfine (HF) levels of opposite parity can be mixed by NSD-PV interactions, and are inherently close in energy.  Accessible laboratory magnetic fields can Zeeman-shift these levels to degeneracy, dramatically enhancing the state mixing.  The matrix element (m.e.) of the NSD-PV interaction can be measured with a Stark-interference technique of demonstrated sensitivity \cite{Nguyen97}.  Use of ground-state molecules leads to enhanced resolution because of their long lifetimes \cite{Fortson93, Nguyen97}.  (Two recent papers also proposed measuring NSD-PV in the ground state HF levels of heavy alkali atoms \cite{Gomez07}.)  The technique is applicable to a wide class of molecules and hence to NSD-PV couplings to the variety of nuclei within them.  We specifically consider diatomic molecules with a single valence electron in a $^2\Sigma$ electronic state.  These are the molecular equivalent of alkali atoms, with a simple, regular structure of rotational/HF levels. This makes it possible to reliably determine the properties of the electronic wavefunction necessary to relate the NSD-PV m.e. to the underlying physics.  Enhancement of NSD-PV using Zeeman-shifted rotational/HF levels of molecules has been proposed before \cite{FK85,KLM}.  We describe for the first time a realistic experimental approach to exploit this enhancement, and emphasize both its generality and its unprecedented sensitivity. 

We illustrate the basic idea of our method using a Hund's case (b) $^2\Sigma$ diatomic molecule \cite{Herzberg}, containing one nucleus ($\mathcal{N}_1$) of spin $I = \frac{1}{2}$ that couples to the valence electron via NSD-PV effects, and a second ($\mathcal{N}_2$) with $I'=0$.  The lowest energy levels are described by the Hamiltonian $H = B_e\mathbf{N}^2 + \gamma\mathbf{N}\cdot\mathbf{S} + b\mathbf{I}\cdot\mathbf{S} + c(\mathbf{I}\cdot\mathbf{n})(\mathbf{S}\cdot\mathbf{n})$, where $N$ is the rotational angular momentum, $S=\frac{1}{2}$ is the electron spin, and $\mathbf{n}$ is a unit vector along the internuclear axis \cite{Childs92} ($\hbar = 1$ throughout).  In most cases of interest, the rotational constant $B_e$ is much larger than the spin-rotation (SR) constant $\gamma$ and the HF constants $b$ and $c$.  Hence $N$ is a good quantum number, with eigenstates of energy $E_N \approx B_e N(N+1)$ and parity $P = (-1)^N$.

We use a magnetic field $\Bvec = \Bsca\hat{z}$ to
Zeeman-shift sublevels of the $N^P = 0^+$ and $1^-$ manifolds of
states to near degeneracy.  Since $B_e \gg \gamma, b, c$, the
magnetic field necessary to bridge the rotational energy $E_1 - E_0
\approx 2 B_e$ is large enough to strongly decouple $\mathbf{S}$ from $\mathbf{I}$ and $\mathbf{N}$.  We write the molecular states in terms of the decoupled basis $|N,m_N\rangle |S,m_S\rangle |I,m_I\rangle$.  In general, the term $c(\mathbf{I}\cdot\mathbf{n})(\mathbf{S}\cdot\mathbf{n})$ in $H$ can lead to significant mixing of such states.  However, for simplicity in the present discussion we assume $c \ll \gamma, b$, so that the decoupled basis states are, to good approximation, the energy eigenstates. The Zeeman effect is dominated by the coupling to $\mathbf{S}$, with
approximate Hamiltonian \cite{KL} $H_Z \cong -g \mu_B \mathbf{S} \cdot
\Bvec$, where $g \cong -2$ and $\mu_B$ is the
Bohr magneton.  Opposite-parity levels $|\psi^+_{\uparrow}(m_N=0, m_I)\rangle \equiv
|0,0\rangle |\frac{1}{2},\frac{1}{2}\rangle |\frac{1}{2},m_I\rangle$ and $|\psi^-_{\downarrow}(m_N',
m_I')\rangle \equiv |1,m_N'\rangle |\frac{1}{2},-\frac{1}{2}\rangle |\frac{1}{2},m_I'\rangle$ are
degenerate under $H+H_Z$ when $\Bsca = \Bsca_0 \approx B_e/\mu_B$.

Pairs of these nearly-degenerate levels can be mixed by NSD-PV
interactions, described by the relativistic Hamiltonian $H'_P =
\kappa' \frac{G_F}{\sqrt{2}} \frac{\hbox{\boldmath{$\alpha$}} \cdot
\mathbf{I}}{I} \delta^3(\mathbf{r})$ \cite{Flambaum80}.  Here $\kappa'$ is a dimensionless number parameterizing the
strength of the NSD-PV interaction, $G_F$ is the Fermi constant, $\hbox{\boldmath{$\alpha$}}$ is the
standard vector of Dirac matrices, and $\mathbf{r}$ is the displacement
of the valence electron from $\mathcal{N}_1$.  Within the
subspace of rotational/HF levels, the effect of $H'_P$ is
described by the effective Hamiltonian $H^\mathrm{eff}_P = \kappa'
W_P C$. Here, $W_P$ characterizes the size of the m.e. of $H'_P$
diagonal in the $^2\Sigma$ wavefunction of the valence
electron (in the molecule-fixed frame); the dimensionless operator $C \equiv (\mathbf{n} \times \mathbf{S}) \cdot \mathbf{I}/I$ encodes the angular momentum dependence of $H^\mathrm{eff}_P$ \cite{FK85}. We seek to determine $\kappa'$, by measuring the NSD-PV m.e.'s $iW(m_N',m_I',m_N, m_I) \equiv \kappa' W_P \langle \psi^-_{\downarrow}(m_N',m_I')|C| \psi^+_{\uparrow}(m_N, m_I)\rangle$. Time-reversal invariance ensures that $iW$ is pure imaginary.

In general, $W_P \propto Z^2$, where $Z$ is the atomic number of $\mathcal{N}_1$ \cite{Bouchiat74,SF}.  The value of $W_P$ can be explicitly calculated, using a semi-empirical method developed earlier by one of us \cite{K85}. Knowledge of $b,c$ and $\gamma$ from standard spectroscopic data \cite{moleculardata} is sufficient to determine the $s$- and $p$-wave components of the valence electron wavefunction near $\mathcal{N}_1$, and hence $W_P$.  The approximations used in this method are expected to give systematic errors of 10-20\%.  This has been explicitly verified in two cases (BaF and YbF) by comparison of the semi-empirical results to sophisticated \textit{ab initio} calculations of $W_P$ \cite{MosyaginJPB31}.  

$C$ is a pseudoscalar, with non-zero m.e.'s $\tilde{C}$
between states with the same value of $m_F \equiv m_N + m_S + m_I$.  In our example, $|\psi^+_{\uparrow}(0, +\frac{1}{2})\rangle$ can mix with $|\psi^-_{\downarrow}(+1, +\frac{1}{2})\rangle$, and $|\psi^+_{\uparrow}(0, -\frac{1}{2})\rangle$ with both $|\psi^-_{\downarrow}(+1,-\frac{1}{2})\rangle$ and $|\psi^-_{\downarrow}(0, +\frac{1}{2})\rangle$.  Level crossings
between the pairs of mixing states occur at slightly different
values of $\Bsca_0$, because of energy differences in the sublevels due
to HF and SR terms in $H$.  The magnetic fields for the various crossings differ by $\Delta \Bsca \ll \Bsca_0$ since $(\gamma,b,c)\ll B_e$.

We calculate $\tilde{C}$ as well as the electric dipole m.e. $d \equiv \langle \psi^+_{\uparrow}| D n_z | \psi^-_{\downarrow} \rangle$, using the definitions $(n_x,n_y,n_z) = (\sin{\theta}\cos{\phi}, \sin{\theta}\sin{\phi}, \cos{\theta})$ and $|N,m_N\rangle = Y_{N}^{m_N}(\theta, \phi)$ (a spherical harmonic).   
Here $\mathbf{D}=D\mathbf{n}$ is the electric dipole moment in the molecular frame.  To first order, $d\! = \! 0$, since $[\mathbf{D},\mathbf{S}]=0$ and $m_S' \neq m_S$.  However, the HF and SR terms in $H$ cause a small mixture of states with different values of $m_S$ into the crossing levels.  The resulting induced values of $d$ can be calculated perturbatively and have typical size $d \sim \eta D$, where $\eta \sim (\gamma,b,c)/B_e \ll 1$.

We have emphasized the essential simplicity of this system by outlining an analytic approach to determining all relevant quantities under some approximations.  However, we have also performed full numerical calculations of energies and m.e.'s for the system.  This reproduces the analytic results in the simple case described here.  It also allows inclusion of additional effects \cite{KL} such as HF structure when $I' \neq 0$; nuclear spin $I > 1/2$ and associated electric quadrupole HF interactions; a $G$-tensor to reflect the anisotropy of the Zeeman interaction in the molecule; Zeeman interactions with nuclear and rotational magnetic moments; etc.  The qualitative behavior of the system is independent of such complicating details.

We measure $iW$ with a Stark-interference method developed for use in atomic Dy \cite{Nguyen97}.  A beam of molecules enters a region of magnetic field $\Bsca \approx \Bsca_0$.  Here the molecules are excited by laser light tuned to resonance with a transition to a short-lived electronic state of definite parity.  Parity selection rules ensure that only one level of the nearly-degenerate ground state pair (say, $|\psi^+_{\uparrow}\rangle$) is excited, and its population is rapidly depleted by optical pumping.  Next, the molecules enter a region of spatially varying electric field $\mathbf{E} = E(z)\hat{z}$, where $E(z)= E_0 \sin{2 \pi N z/L}$ for $0\!<\!z\!<\!L$ ($N$ is an integer). In the rest frame of molecules with velocity $\mathbf{v}\!=\!v\hat{z}$, this is a time-varying field $E(t\!=\!z/v)=E_0 \sin{\omega t}$, where $\omega = 2 \pi N v/L$.  The Hamiltonian $H_{\pm}$ for the two-level system of near-degenerate states can be written as
\begin{equation}
H_{\pm} = \TwoByTwo{0}{iW+dE}{-iW+dE}{\Delta},
\end{equation}
where $\Delta$ is the small $\Bsca$-dependent detuning from exact degeneracy under $H+H_Z$.  The wavefunction is 
\begin{equation}
|\psi(t)\rangle = c_+(t)|\psi^+_{\uparrow}\rangle + e^{-i\Delta t}c_-(t)|\psi^-_{\downarrow}\rangle \equiv \TwoColumn{c_+}{c_-},
\end{equation}
with $c_+(0) = 0$ due to the optical pumping.  Assuming $W \ll (dE_0, \Delta) \ll \omega$, the Schr\"odinger equation yields
\begin{equation}
\begin{split}
c_+(t)  &=  -2 i e^{-\frac{i\Delta t}{2}}
\times  \left[\cos\left(\frac{\Delta t}{2}\right)\frac{dE_0}{\omega}\sin^2\left(\frac{\omega t}{2}\right)\right. \nonumber\\
&\left. + i\sin\left(\frac{\Delta t}{2}\right)\left\{\frac{W}{\Delta}+\frac{dE_0}{\omega}\cos^2\left(\frac{\omega t}{2}\right)\right\} \right]. \label{solnSchrEqnwitht}
\end{split}
\end{equation}
At the end of the electric field region, $t=T\equiv L/v$ and $\omega T = N \pi$, \emph{regardless of} v. Here $c_+$ has the final value
\begin{equation}
c_+(T) = 2 e^{-\frac{i\Delta T}{2}}\sin\!\left(\frac{\Delta T }{2}\right)\left\{\frac{W}{\Delta}+\frac{d E_0}{\omega}\right\}.
\end{equation}
Next, the population of the initially depleted state is measured, e.g. by collecting laser-induced fluorescence from this state.  This yields a signal $S$:
\begin{equation}
\begin{split}
S &= N_0 \left|c_+(T)\right|^2 \nonumber\\
&\cong 4N_0{\rm sin}^2\!\left(\frac{\Delta T}{2}\right)\left[2 \frac{W}{\Delta}\frac{d E_0}{\omega} + \left(\frac{d E_0}{\omega}\right)^2\right], \label{signal}
\end{split}
\end{equation}
where $N_0$ is the number of molecules that would be detected in the absence of the optical pumping laser and electric field. We define the PV asymmetry $\mathcal{A}$ formed by reversal of the electric field $E_0$ as:
\begin{equation}
\mathcal{A} \equiv \frac{S(+E_0) - S(-E_0)}{S(+E_0) + S(-E_0)}= 2\frac{W}{\Delta}\frac{\omega}{d E_0}. 
\end{equation}

Formally, $\mathcal{A}$ diverges for $\Delta = 0$.  However, even in ideal conditions the uncertainty in $W$ is limited by shot noise to $\delta W = (W \mathcal{A} \sqrt{2S})^{-1} = \Delta/(4\sqrt{2N_0}\sin{\left(\frac{\Delta T}{2}\right)})$, which reaches a minimum value $\delta W_{\mathrm{min}} = 1/(2\sqrt{2N_0}T)$ as $\Delta\rightarrow 0$.  In addition, under realistic conditions the magnetic field $\Bsca$ will not be perfectly homogeneous, so that the values of $\Delta$ must be averaged over the ensemble of molecules.  If $\Delta$ is described by a distibution of mean value $\Delta_0$ and r.m.s. deviation $\Gamma$, then for small detunings $(\Delta, \Gamma \ll 2/T)$ the ensemble-averaged asymmetry is $\left\langle\mathcal{A}\right\rangle = 2\frac{W}{\Delta_0}\frac{\omega}{d E_0}\frac{\Delta_0^2}{\Delta_0^2+\Gamma^2}$, which vanishes at $\Delta_0 = 0$.  Near-minimum uncertainty $\delta W \approx \delta W_{\mathrm{min}}$ is then obtained when $\Gamma$ and $\Delta_0$ satisfy $\Gamma \lesssim \Delta_0 \lesssim 2/T$.

We consider molecules with rotational constant $B_e \lesssim 2 \pi \times 30$ GHz, which require magnetic fields $\Bsca_0 \lesssim 2$ T for level crossings.  Nearly all atoms from periodic table groups 2,13,16, and 17, plus some from groups 1,3,4,14, and 15 appear in $^2\Sigma$ diatomic molecules satisfying this criterion \cite{moleculardata}.  All have laser-accessible electronic transitions for optical pumping.  A standard technique for producing beams of $^2\Sigma$ free radicals \cite{Steimle,Tarbutt} can yield molecular velocities $v \lesssim 5\times10^4$ cm/s.  With an electric field region $L \sim 5$ cm, $\delta W \approx \delta W_{\mathrm{min}}$ can be achieved if $\Gamma \lesssim 2 \pi \times 3$ kHz, corresponding to an r.m.s. deviation in the magnetic field $\delta \Bsca_0 / \Bsca_0 \lesssim \Gamma/B_e \sim 10^{-7}$.  The desired field strength and homogeneity are available with commercial magnetic resonance imaging magnets.  Molecular beam fluxes $F \sim 10^{10}$/sr/s in the $N=0$ state have been achieved for $^2\Sigma$ free radical species such as YbF and CaF \cite{Tarbutt}.  $N_0$ depends on the solid angle of beam intercepted ($\Omega$), the detection efficiency ($\eta$), and the fraction of population in a single Zeeman sublevel of the desired isotope ($f$).  With realistic estimates $\Omega \sim 10^{-5}$ and $\eta \sim 5 \times 10^{-2}$, and a typical value $f \sim 10^{-2}$, we expect $N_0 =f\eta\Omega F \sim 50$ mol/s and hence $\delta W \sim 2 \pi \times 80$ Hz/$\sqrt{\mathrm{\tau}}$, where $\tau$ is the total integration time in seconds.  

From $W$ (measured) and $W_P$ (calculated), we determine $\kappa' = \kappa_2' + \kappa_a' + \kappa_Q'$ \cite{FlamAnapoleDDH}.  Here, $\kappa_2'$ arises from the $V_{\rm e}A_{\rm n}$ term in electron-nucleus Z$^0$-exchange;  $\kappa_a'$ from the nuclear anapole moment; and $\kappa_Q'$ from the coherent effect of $Q_{\rm W}$ and the magnetic HF interaction \cite{KappaQ}.  $\kappa_Q'$ is small compared to the other terms, and well-understood; we ignore it henceforth.  In any measurement on a given nucleus, the effects of $\kappa_a'$ and $\kappa_2'$ are indistinguishable.  However, they can be separated by measurements over a range of nuclei \cite{BPZFP}, since $\kappa_2'$ is independent of the nuclear mass $A$ while $\kappa_a' \propto A^{2/3}$. Thus, in heavy nuclei the anapole moment dominates the NSD-PV effect, while in light nuclei tree-level Z$^0$ exchange is primary.  
\begin{table*}
\begin{tabular}{|c|c|c|c|c|c|c|c|c|c|c|c|c|c|c|c|c}
\hline
Nucleus&$I$&$\nu$&$\ell$&n.a.&$100\!\times\!\kappa_a'$&$100\!\times\!\kappa_2'$&Species&$B_e$&$\Bsca_0^{(\mathrm{m})}$&$W_P$&$\tilde{C}^{(\mathrm{m})}$&$W^{(\mathrm{m})}$&$f$&$D$&$d^{(\mathrm{m})}$\\
       &   &     &      & (\%) &              &              &       &(MHz)& (T)          &(Hz) &  &(Hz)&(\%) &(Debye)&(kHz$\cdot$cm/V)\\
\hline
$^{87}$Sr$_{38} $&9/2&N&4&7.0 &-3.6         &-5.0&     SrF&   7515       &0.62           &65 &-0.40&2.2&0.2&$3.5$&-4.6\\
\hline
$^{91}$Zr$_{40}$&5/2&N&2&11.2 &-3.5         &-5.0&     ZrN&   14468      &1.20            &99 &-0.40& 3.4&0.3&$\sim\!4$&$\sim\! 1.2$\\
\hline
$^{137}$Ba$_{56}$&3/2&N&2&11.2&+4.2         &+3.0&     BaF&   6480       &0.32           &164&-0.44&-5.2&0.7&$3.2$&-3.0\\
\hline
$^{171}$Yb$_{70}$&1/2&N&1&14.3&+4.1         &+1.7&     YbF&   7246       &0.33          &729&-0.52&-22&1.8&$3.9$&1.5\\
\hline
$^{27}$Al$_{13} $&5/2&P&2&100 &-11.2        &+5.0&     AlS&   8369       &0.52           &10 &-0.42&0.3&8&$3.6$&2.5\\
\hline
$^{69}$Ga$_{31} $&3/2&P&1&60.1&-19.6        &+5.0&     GaO&   8217       &0.49           &61 &-0.43&3.8&8&$\sim\!4$&$\sim\! -33$\\
\hline
$^{81}$Br$_{35} $&3/2&P&1&49.3&-21.8        &+5.0&    MgBr&   4944       &0.34           &18 &-0.42&1.3&6
&$\sim\!4$&$\sim\! -6.3$\\
\hline
$^{139}$La$_{57}$&7/2&P&4&99.9&+34.7        &-3.9&     LaO&  10578        &0.25           &222 &-0.43&-29&6&$3.2$&0.6\\
\hline
\end{tabular}
\caption{Data relevant to the proposed measurements, for a sample of nuclei in molecules where all necessary spectroscopic data is available.  Most parameters are defined in the text.  Superscript $(\mathrm{m})$ indicates the value at the level-crossing with the maximum value of $m_F$.  Nuclear shell-model quantum numbers are from Ref. \cite{Kopfermann}; molecular data from Ref. \cite{moleculardata}. \label{tab:01}}
\end{table*}

The values of $\kappa_a'$ and $\kappa_2'$ can be estimated \cite{FlamAnapoleDDH} from a simple model of the nucleus, consisting of a single valence nucleon [$\nu = N (P)$ for a neutron (proton)] with orbital angular momentum $\ell$ around a uniform core:
\begin{equation}
\kappa_a' \approx 1.0\times 10^{-3} g_\nu \mu_\nu A^{2/3} \frac{K}{I+1}; \kappa_2' = C_{2\nu} \frac{1/2 - K}{I+1}.
\end{equation}
Here, $K\!=\!(I\!+\!1/2)(-1)^{I+1/2-\ell}$, $g_\nu$ describes the strength of the hadronic PV interaction between $\nu$ and the core, and $\mu_\nu$ is the nucleon magnetic moment in nuclear magnetons.  In the SM, $C_{2P,N}$ are given at tree level by \cite{Reviews}
\begin{equation}
C_{2{\rm P}} = - C_{2{\rm N}} = \lambda\left(1-4\sin^2\theta_{\rm W}\right)/2 \approx 0.05,
\end{equation}
where $\theta_{\rm W}$ is the weak mixing angle and $\lambda \approx 1.25$. The values of $g_{P,N}$ can be related \cite{FlamAnapoleDDH,HaxWie} to a set of parameters describing low-energy hadronic PV interactions, e.g. the ``DDH'' set \cite{DDH}.  Based on current knowledge, $g_P \approx 4-6$ and $g_N \approx -(0.2-1)$ are roughly expected\cite{FlamAnapoleDDH,HaxWie}; we take $g_P = 5$ and $g_N = -1$ for concreteness.  Table~\ref{tab:01} shows a sample of experimentally accessible cases. For each nucleus listed, we project that $\kappa'$ can be measured to $\sim 10\%$ precision in an integration time of $<1$ week.

The proposed technique should also provide excellent control over systematic errors.  The asymmetry $\mathcal{A}$ resulting from reversal of $E_0$ is also odd in $\Delta$ and in the sign of $\Bsca_0$. (It corresponds to the P-odd invariant $\frac{d\mathbf{E}}{dt}\cdot\left(\Bvec-\Bvec_0\right)$ \cite{Nguyen97}.)  From Ref. \cite{Nguyen97}, it is known that the most troubling systematic errors (surviving all three reversals) arise from stray electric fields in combination with magnetic field gradients.  The suppressed values of $d$ mean that, even before reversal of $\Delta$ and $\Bvec$, macroscopic stray fields $\gtrsim \! 0.1$ mV/cm are needed to mimic the effect of $iW$, for all nuclei considered.  In addition, we find that the complex angular momentum dependence of the relevant operators makes the ratio $\tilde{C}/d$ vary widely (but deterministically) in magnitude and sign between nearby level crossings in the same molecule.  This provides an additional, powerful test for systematics.

A set of values for $\kappa'$ from many nuclei will have impact in both nuclear and particle physics.  A global fit to all experimental data on hadronic PV at present yields large error bars and is internally inconsistent \cite{FlamAnapoleDDH,HaxWie}.  Measurements in several nuclei, with both $\nu=N$ and $\nu=P$, should be sufficient to determine the hadronic PV parameters responsible for $\kappa_a'$, to moderate accuracy.  The limiting uncertainty is likely to arise from the calculations of nuclear structure needed to relate the anapole moment to the DDH parameters. At present, accuracy of $\sim\! 30\%$ appears reasonable, based on the spread in calculated values \cite{DetailedAnapoleCalcs,HaxWie,FlamAnapoleDDH}.  
After several measurements in heavier nuclei, the residual effect of $\kappa_a'$ in the lightest nuclei could hence be subtracted away with $\sim 30\%$ uncertainty, and $C_2$ determined with relative uncertainty $\delta C_2/C_2 \approx 0.3(\kappa_a'/\kappa_2')$.  This is most favorable for $C_{2{\rm N}}$, since $\kappa_a'$ is expected to be suppressed for odd-N nuclei by the small value of $g_N$.  We project $\delta C_{2{\rm N}} / C_{2{\rm N}} \lesssim 20\%$ is possible.  $C_{2{\rm P,N}}$ can be written as linear combinations of $C_{2{\rm u,d}}$; e.g., $C_{2{\rm N}} \cong 0.85C_{2{\rm d}} - 0.40C_{2{\rm u}}$ including SU(3) and radiative corrections \cite{CahnKaneMarSir}.  $C_{2{\rm P,N}}$ have never been measured directly, and other linear combinations of $C_{2{\rm u,d}}$ have experimental uncertainties of 70-300\% \cite{PDG06, Beise}.  A measurement of $C_{2{\rm N}}$ would be complementary to a planned precise measurement of $2C_{2{\rm u}}-C_{2{\rm d}}$ \cite{Reimer04}.

In summary, we propose a new technique for measuring NSD-PV in a broad range of nuclei.  We plan to experimentally implement the method using $^{137}$BaF as the first case.  In the longer term, these ideas might be significantly extended.  The sensitivity might be improved with new molecular beam sources that promise dramatically higher $F$ at lower $v$ \cite{DoyleBeams}.  Alternatively, it might be possible to increase $T$ using trapped molecular ions \cite{Stutz04}. Such improvements, plus more spectral data for similar molecular species, could widen the list of accessible nuclei.  For light molecules and nuclei, \textit{ab initio} electronic and nuclear structure calculations may be possible at accuracies better than those envisioned here.  Ultimately, the method might extend to direct measurement of $\kappa'$ for $^1$H and $^2$H.  For diatomics, the low mass of H makes $B_e$ and hence $\Bsca_0$ prohibitively large.  However, use of simple triatomic species such as HSiO may circumvent this difficulty.

This work was supported by NSF Grant 0457039. 

\end{document}